\documentclass[final,5p,times,twocolumn]{elsarticle}

\usepackage{amssymb}
\usepackage{lipsum}
\usepackage{booktabs} 

\usepackage{cleveref}

\setcitestyle{square}  

\usepackage{lineno}

\journal{Nuclear Instruments and Methods in Physics Research A}

\begin{document}

\begin{frontmatter}



\title{Distillation and Stripping purification plants for JUNO liquid scintillator}

\author[b]{C.\,Landini\corref{cor1}}\ead{cecilia.landini@mi.infn.it}
\author[b]{M.\,Beretta}
\author[b]{P.\,Lombardi}
\author[b]{A. Brigatti}
\author[h,i]{M. Montuschi}
\author[b]{S. Parmeggiano}
\author[b]{G. Ranucci}

\author[b]{V. Antonelli}
\author[b]{D. Basilico}
\author[b]{B. Caccianiga}
\author[b]{M.G. Giammarchi}
\author[b]{L. Miramonti}
\author[b]{E. Percalli}
\author[b]{A.C. Re}
\author[b]{P. Saggese}
\author[b]{M.D.C. Torri}

\author[a]{S. Aiello}
\author[a]{G. Andronico}
\author[e]{A. Barresi}
\author[k]{A. Bergnoli}
\author[e]{M. Borghesi}
\author[c]{R. Brugnera}
\author[a]{R. Bruno}
\author[d]{A. Budano}
\author[j]{A. Cammi}
\author[c]{V. Cerrone}
\author[a]{R. Caruso}
\author[e]{D. Chiesa}
\author[f]{C. Clementi}
\author[k]{S. Dusini}
\author[d]{A. Fabbri}
\author[g]{G. Felici}
\author[c]{A. Garfagnini}
\author[a]{N. Giudice}
\author[c]{A. Gavrikov}
\author[c]{M. Grassi}
\author[c]{R.M. Guizzetti}
\author[a]{N. Guardone}
\author[c]{B. Jelmini}
\author[c]{L. Lastrucci}
\author[k]{I. Lippi}
\author[j]{L. Loi}
\author[a]{C. Lombardo}
\author[h,i]{F. Mantovani}
\author[d]{S.M. Mari}
\author[g]{A. Martini}
\author[e]{M. Nastasi}
\author[d]{D. Orestano}
\author[f]{F. Ortica}
\author[g]{A. Paoloni}
\author[d]{F. Petrucci}
\author[e]{E. Previtali}
\author[k]{M. Redchuck}
\author[h,i]{B. Ricci}
\author[f]{A. Romani}
\author[a]{G. Sava}
\author[c]{A. Serafini}
\author[c]{C. Sirignano}
\author[e]{M. Sisti}
\author[k]{L. Stanco}
\author[d]{E. Stanescu Farilla}
\author[h,i]{V. Strati}
\author[c]{A. Triossi}
\author[a]{C. Tuve'}
\author[d]{C. Venettacci}
\author[a]{G. Verde}
\author[g]{L. Votano}

\cortext[cor1]{Corresponding author}
\address[b]{INFN, Sezione di Milano e Università degli Studi di Milano, Dipartimento di Fisica, Italy}
\address[h]{INFN, Sezione di Ferrara, Italy}
\address[i]{Università degli Studi di Ferrara, Dipartimento di Fisica e Scienze della Terra, Italy}
\address[a]{INFN, Sezione di Catania e Università di Catania, Dipartimento di Fisica e Astronomia, Italy}
\address[k]{INFN, sezione di Padova, Italy}
\address[c]{INFN, sezione di Padova e Università di Padova, Dipartimento di Fisica e Astronomia, Italy}
\address[d]{INFN, sezione di Roma Tre e Università degli Studi Roma Tre, Dipartimento di Fisica e Matematica, Italy}
\address[e]{INFN, Sezione di Milano Bicocca e Dipartimento di Fisica Università di Milano Bicocca, Italy}
\address[f]{INFN, Sezione di Perugia e Università degli Studi di Perugia, Dipartimento di Chimica, Biologia e Biotecnologie, Italy}
\address[g]{Laboratori Nazionali dell’INFN di Frascati, Italy}
\address[j]{INFN, Sezione di Milano Bicocca e Dipartimento di Energetica, Politecnico di Milano, Italy}

\begin{abstract}
The optical and radiochemical purification of the scintillating liquid, which will fill the central detector of the JUNO experiment, plays a crucial role in achieving its scientific goals. Given its gigantic mass and dimensions and an unprecedented target value of about 3$\%$/$\sqrt{E(MeV)}$ in energy resolution, JUNO has set severe requirements on the parameters of its scintillator, such as attenuation length (L\textsubscript{at}\textgreater20 m at 430 nm), transparency, light yield, and content of radioactive contaminants (\textsuperscript{238}U,\textsuperscript{232}Th\textless10\textsuperscript{-15} g/g). To accomplish these needs, the scintillator will be processed using several purification methods, including distillation in partial vacuum and gas stripping, which are performed in two large scale plants installed at the JUNO site.

In this paper, layout, operating principles, and technical aspects which have driven the design and construction of the distillation and gas stripping plants are reviewed. The distillation is effective in enhancing the optical properties and removing heavy radio-impurities (\textsuperscript{238}U,\textsuperscript{232}Th, \textsuperscript{40}K), while the stripping process exploits pure water steam and high-purity nitrogen to extract gaseous contaminants (\textsuperscript{222}Rn, \textsuperscript{39}Ar, \textsuperscript{85}Kr, O\textsubscript{2}) from the scintillator. The plant operating parameters have been tuned during the recent commissioning phase at the JUNO site and several QA/QC measurements and tests have been performed to evaluate the performances of the plants. Some preliminary results on the efficiency of these purification processes will be shown.

\end{abstract}




\end{frontmatter}





\section{Introduction}
\label{Ch intro}
In the research field of neutrino physics, liquid detectors have been widely used to investigate the properties of these weakly interacting particles. Organic scintillators have been chosen as detection medium by several experiments, such as Borexino~\cite{Borex}, KamLAND~\cite{KamLand}, Daya Bay~\cite{DayaBAy}, RENO~\cite{RENO}, JUNO~\cite{JUNO_PPNP} and SNO+~\cite{SNO}, and the size and masses involved as detector target have kept on growing over the years.

Due to their homogeneity, flexible handling, affordability, availability in large quantities and possibility of different purification methods, liquid scintillators represent an optimal solution. So far, Borexino and KamLAND above all managed to achieve extraordinary results for radiopurity levels even in the kiloton scale.

JUNO (Jiangmen Underground Neutrino Observatory) is a new-generation liquid-scintillator reactor antineutrino experiment, currently under construction in an underground laboratory (700 m vertical overburden) near Kaiping, Guandong province, in South China. The primary goal is the determination of the neutrino mass ordering at 3$\sigma$ level in 6 years, by detecting reactor antineutrinos from two nuclear power plants, located approximately 53 km away. The gigantic central detector (CD) is composed by a transparent acrylic sphere of 35.4 m in diameter, designed to contain about 20.000 ton of liquid scintillator (LS). 17.612 large (20") photomultipliers and 25.600 small (3") photomultipliers (PMTs) installed around the acrylic vessel will detect the scintillation light produced inside, providing a large total coverage of about 78$\%$. These features are designed to reach an unprecedented energy resolution for a liquid scintillator detector of about $\sim$3$\%$/$\sqrt{E(MeV)}$. The CD is immersed into a high-purity water pool, aiming to shield the active region from the natural radioactivity of the external environment and acting also as a muon veto detector, together with the muon veto top tracker~\cite{TopTracker}, to reject the cosmic rays background.

To accomplish the extensive scientific program of JUNO~\cite{OscillParam}~\cite{FeasibilityB8}~\cite{CoreColl}~\cite{PDecay}~\cite{GalacticHAlo}, a high sensitivity and an extremely low background~\cite{RadBackground} are mandatory. For this purpose, one of the crucial tasks is the purification of the liquid scintillator, which is pivotal to achieve the optical and radiopurity levels required for the scintillator. Indeed, if radioisotopes are present inside the scintillating mixture, their signal will be detected, superimposing to the real neutrino signal, while optical impurities could spoil its transparency and affect the light propagation and detection, and hence the overall energy resolution. Several chemical techniques are available to purify liquid scintillators, but the challenging needs of neutrino physics frontier research imply special attentions and a strong push forward in technological and engineering solutions beyond the state-of-the-art.

In the following, the purification strategy of JUNO will be presented, paying particular attention to the distillation and gas stripping purification processes.

The layout of this paper manages as follows. Section~\ref{Ch LS} briefly introduce the Juno liquid scintillator and its purification strategy. The description of the distillation and stripping plants is shown and discussed in section~\ref{Ch DistStr}. Finally, plants commissioning and preliminary purification results are given in section~\ref{Ch Commissioning}.


\section{Juno liquid scintillator}
\label{Ch LS}
The scintillator of JUNO is based on organic solvent, the linear alkyl benzene (LAB), whose formula can be written as C\textsubscript{6}H\textsubscript{5}C\textsubscript{n}H\textsubscript{2n+1}, with n typically between 10 and 14. The fundamental part of the molecule is the benzene ring, which can be excited by ionizing radiation, thus assuring the scintillating feature. Given the low costs and possibility of huge scale mass production, high transparency and compatibility with acrylic, good properties in terms of safety and health hazards, the LAB represents an optimal choice as solvent. As primary fluor, the 2,5-Diphenyloxazole (PPO) is added to it, to enhance the scintillation, while the 1,4-Bis(2-methylstyryl)benzene (bis-MSB) has been selected as wavelength shifter, to further reduce the self-absorption and optimize the coupling with the PMTs in the wavelength region of their highest sensitivity, i.e. around 430 nm.

The scintillating mixture was optimized with a dedicated test campaign at DayaBay laboratories~\cite{LightY}, where different concentrations of PPO and bis-MSB were added and studied. The final JUNO LS recipe has been set to have 2,5 g/L of PPO and 3 mg/L of bis-MSB.

The experimental goals of JUNO impose several requirements on the LS, mainly in terms of optical and radiopurity features. Since a low background is mandatory for JUNO to unravel rare neutrino events, the content of radio-impurities and radioisotopes dissolved in the scintillator must be minimized, to prevent undesired signals. Typically, the main sources of contamination, that could be naturally present in the scintillator or introduced from the external environment (dust, air, residual microparticles washed off and/or emanated from surfaces contacted with liquid), are \textsuperscript{238}U, \textsuperscript{232}Th, \textsuperscript{40}K as heavy impurities and \textsuperscript{222}Rn, \textsuperscript{85}Kr and \textsuperscript{39}Ar as gaseous impurities. The minimum radiopurity levels in JUNO are settled for the detection of antineutrinos from the nearby nuclear plants, consisting in \textsuperscript{238}U,\textsuperscript{232}Th\textless10\textsuperscript{-15}g/g and \textsuperscript{40}K\textless10\textsuperscript{-16}g/g. Instead, for solar neutrino studies, the levels must be lowered at 10\textsuperscript{-16}g/g, or even below for better precision measurements. Even at 10\textsuperscript{-16}g/g, we foresee to improve the Borexino’s measurements for \textsuperscript{7}Be neutrinos in 1 year of data taking and for neutrinos from CNO cycle in 6 years of data taking~\cite{Solar}, which is the same acquisition time needed to determine the neutrino mass ordering at 3$\sigma$ level. In general, the optimal level for \textsuperscript{238}U/\textsuperscript{232}Th/\textsuperscript{40}K impurities would be less than 10\textsuperscript{-17} g/g, established as final target value in JUNO. More details about JUNO radiopurity requirements are reported in Table~\ref{tab:radio_requirements}.

Concerning the optical requirements, the transparency and the attenuation length must be carefully controlled to maintain an optimal light propagation and thus a good energy resolution. Given the huge dimensions of the central detector (35,4 m in diameter), the optical path of the scintillation light before reaching the PMTs outside could be very long. Hence, the requirement on the attenuation length of the final scintillator mixture is fixed to be L\textsubscript{at}\textgreater20 m at 430 nm. The light yield is expected to be around 1.500 p.e./MeV, leading to an unprecedented energy resolution of 3$\%$ at 1 MeV. The energy response linearity and the long-term stability with ageing are other important features to be considered.

To satisfy all the stringent requirements of JUNO LS, a dedicated procedure of purification in several steps has been studied and implemented for the scintillator and is described in the next sub-section.

\begin{table*}[t]
    \centering
    \begin{tabular}{c|c|c|c}
        \toprule
        Radioisotope & Contamination source & Typical value & JUNO minimum requirements \\
        \midrule
        \textsuperscript{222}Rn & Air, emanation from material & \textless100 Bq/m\textsuperscript{3} & \textless5 mBq/m\textsuperscript{3} \\ 
        \textsuperscript{238}U & Dust suspended in liquid & $\sim$10\textsuperscript{-6} g/g & \textless10\textsuperscript{-15} g/g \\ 
        \textsuperscript{232}Th & Dust suspended in liquid & $\sim$10\textsuperscript{-5} g/g & \textless10\textsuperscript{-15} g/g \\ 
        \textsuperscript{40}K & Dust suspended in liquid, PPO & $\sim$10\textsuperscript{-6} g/g & \textless10\textsuperscript{-16} g/g \\ 
        \textsuperscript{39}Ar & Air & $\sim$1 Bq/m\textsuperscript{3} & \textless50 $\mu$Bq/m\textsuperscript{3} \\ 
        \textsuperscript{85}Kr & Air & $\sim$1 Bq/m\textsuperscript{3} & \textless50 $\mu$Bq/m\textsuperscript{3} \\
        \bottomrule
    \end{tabular}
    \caption{List of the main radio-contaminants that can be present in JUNO liquid scintillator, the relative contamination sources and their typical values. The minimum radiopurity levels for JUNO LS, required for the antineutrino reactor program, are reported in the last column.}
    \label{tab:radio_requirements}
\end{table*}

\subsection{Purification strategy for Juno LS}
\label{Ch purif_strategy}
As done in other neutrino experiments based on liquid detectors, a purification strategy with different chemical techniques is crucial to suppress the background deriving from radio-contaminants present in the scintillator itself and improve its purity up to the required standard. In JUNO, the LS is processed by a sequence of 5 purification systems:

\begin{itemize}
\item filtration of raw LAB through Al\textsubscript{2}O\textsubscript{3} (alumina) powder, in order to improve its optical properties, increase the attenuation length and smoothen the absorption spectrum~\cite{Boxiang};
\item distillation of LAB in partial vacuum, useful to remove heavy and high-boiling impurities, such as \textsuperscript{238}U, \textsuperscript{232}Th and \textsuperscript{40}K, and further enhance the optical properties;
\item acid washing of the Master Solution, created in batches with a high concentration of PPO and bis-MSB in LAB, and then dilution until the JUNO recipe;
\item water extraction of LS, which is effective to remove polar contaminants and ions that may contain \textsuperscript{238}U, \textsuperscript{232}Th and \textsuperscript{40}K isotopes~\cite{ZioFang};
\item gas stripping of LS, in order to extract radioactive gases like \textsuperscript{222}Rn, \textsuperscript{39}Ar, \textsuperscript{85}Kr.
\end{itemize}

For each of these steps, a dedicated large-scale plant has been built and installed at JUNO experimental site. After completing the first 3 processes over ground, the LS is sent to the underground laboratory by a DN50 stainless steel pipe through the 1,5 km slope tunnel, for the last 2 steps of the sequence. In case of unqualified samples or if the LS already filled in the CD needs to be re-purified along JUNO’s lifetime, it can be sent back and re-processed underground by water extraction and stripping plants, the only two techniques allowed after that PPO and bis-MSB have already been added to the scintillating mixture.

\section{Distillation and Stripping plants overview}
\label{Ch DistStr}
Among the complex purification procedure for JUNO LS described in section~\ref{Ch purif_strategy}, this paper focuses on the details and performances of distillation and stripping purification plants (steps 2 and 5 of the sequence, respectively), which were entirely designed and build in Italy in collaboration with Polaris Engineering company, Misinto (MB), Italy. The full-scale plant design was optimized and finalized with the experience earned during the manufacturing and commissioning of two small-scale pilot plants, installed at Daya Bay laboratories~\cite{Pilota}.

All the plants have been realized in compliance with both Chinese and European standards and safety rules.

\subsection{Pilot Plants}
\label{Ch pilots}
In order to study the feasibility and the purification efficiency of these techniques on a LAB based liquid scintillator, dedicated small-scale pilot plants were realized and installed at Daya Bay Neutrino laboratory, near Shenzhen, in China. For distillation and stripping processes, the research and design phase begun in 2014, involving INFN (Istituto Nazionale di Fisica Nucleare) and Polaris Engineering in close cooperation. Given the limited budget, the dimensions of each pilot plant were relatively small, fitting entirely in one 2.15 m × 2.4 m × 7 m skid, also for convenience of transportation and handling, and with a maximum flow rate of only 100 kg/h for the purified stream.

After the construction was completed in 2016, the pilot plants were shipped to China by sea and installed at Daya Bay. The commissioning and intensive test campaign were performed in 2017-2018.

The idea was to test the purification techniques and develop a deep know-how that would have been fundamental for the optimization of the process parameters in the full-scale plants. All the details of these pilot plants can be found in~\cite{Pilota}.

The essential components necessary for the two processes were implemented, but the level of plant automation was reduced to few automatic valves and controllers, for cost-effectiveness. This aspect has been robustly upgraded in the final plants, thus assuring a safe and automatized system able to run stably and independently in nominal conditions and in fail-safe conditions in case of emergencies.

For the circulation of the scintillator, the membrane dosing pumps of the pilot plants have been substituted in the full-scale plants by magnetic driven centrifugal pumps, more reliable and suitable for higher flow rates, and an automatic valve to control the flux. Different types of sensors were tested, especially to measure the level and pressure inside the distillation and stripping columns, in the harsh conditions of partial vacuum. The number of vacuum pumps has been increased from 1 to 4, both to satisfy the required pumping speed and for redundancy, thus also allowing maintenance operations without stopping the plant. Several other features have been later improved on the basis of the knowledge acquired in this preparatory phase.

The pilot plants campaign has been successfully concluded with excellent results about a real effectiveness in enhancing the optical and radiopurity features of the JUNO LAB-based scintillator and laying the foundations for the development of the full-scale purification plants, in a profitable and lasting collaboration with Polaris company.

\subsection{Distillation plant}
\label{Ch Dist}
The main goal of the distillation plant is the removal of the heaviest impurities from LAB, like heavy metals and particulate that could contain \textsuperscript{238}U, \textsuperscript{232}Th and \textsuperscript{40}K radioisotopes. This process has been proven effective also to enhance the optical quality of the solvent, mainly the absorbance and the attenuation length in the 340-500 nm range of wavelengths.

\subsubsection{Working principle of the distillation column}
\label{Ch Dist_work}
The fractional distillation is a chemical technique widely used to separate or extract from a liquid mixture some of its components, called fractions. Taking advantage of the different boiling points of the fractions, the mixture is heated up until selective boiling and subsequent condensation of the desired component occurs. The most volatile elements concentrate in higher percentage in the evaporated phase, while the less volatile ones remain in the liquid.

In our plant, the distillation process is carried out in a 7 m-high, 2000 mm-wide custom-made distillation column. In the bottom part, a reboiler, heat up by hot oil, is directly connected to the column to boil the LAB and to produce purified vapours, which are then collected and liquified at the top of the column by a condenser.

The column is equipped with 6 sieve trays (dual flow trays) with $\sim$3500 holes (12 mm in diameter), where a layer of liquid can build-up, ensuring an intimate contact with the upward stream of vapours. The purification process is driven by both heat and mass transfers between the gas and liquid phases at each tray, thus forming multiple equilibrium conditions along the height of the column. The vapor and liquid fluxes should be limited to the design operating range, so that a suitable contact surface and time at each stage can be ensured.

The distillation of LAB is performed at 210-220°C in partial vacuum (60 mbar at column bottom), in order to reduce the  boiling temperature, avoiding  risks of thermal degradation and increasing the separation capability with respect to heavy impurities.

The vacuum level inside the column is kept constant by 4 vacuum pumps (VP) connected in the upper part, after the condenser. Pressure is the key operating parameter that drives all the others inside the column: by setting the pressure values at the bottom (60 mbar at the reboiler) and top (5 mbar at the condenser) parts, the first one determines the boiling temperature of the liquid, which is constant during phase transitions, while the latter is adjusted by VPs through an automatic regulating valve. The pressure difference between them corresponds to the total height of liquid accumulated on the trays: considering 55 mbar of pressure difference and a LAB density of 720 kg/m\textsuperscript{3} at 200°C average temperature, the height is approximately 78 cm, thus implying a liquid layer of $\sim$13 cm on each tray. So, in this design, the liquid is held on the perforated tray by the pressure of the lower stage, in a dynamic equilibrium between evaporation and condensation. If pressure changes abruptly, the layers could break and the liquid would fall to the bottom of the column. Since the layer height on the trays affects the global purification efficiency of the distillation, the pressure values inside the column must be carefully controlled to avoid any destabilization of it.

\begin{figure}
\centering
\includegraphics[height=6cm]{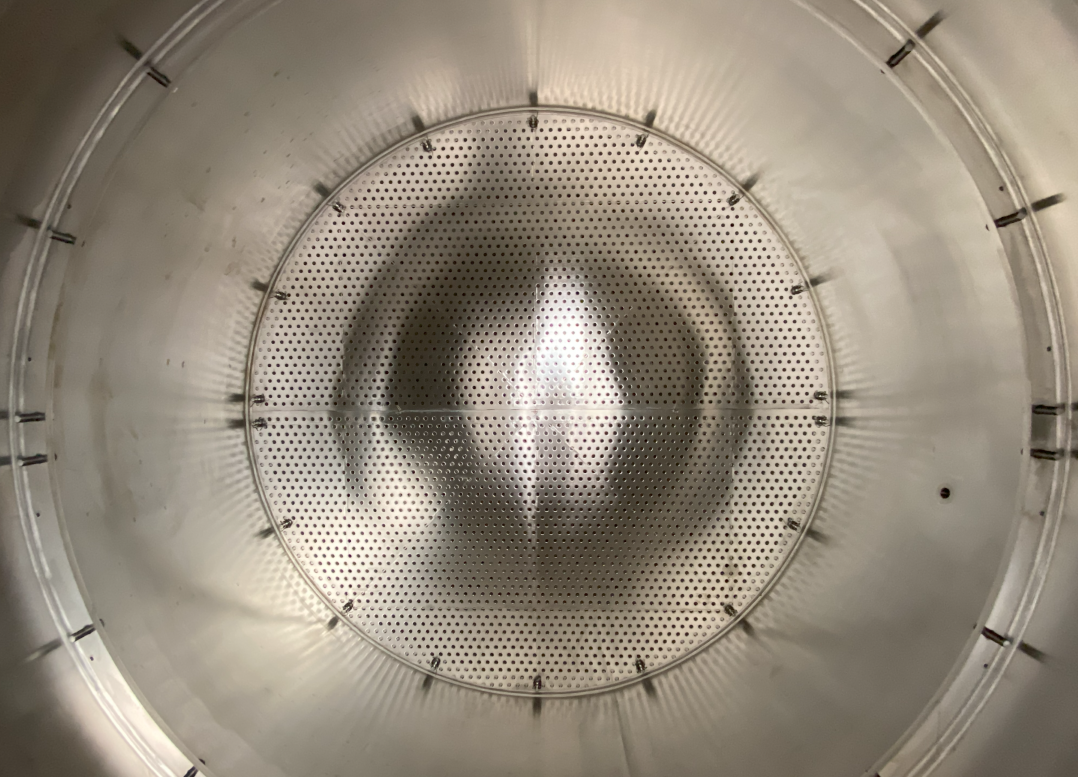}     
\caption{Installation of sieve trays inside the distillation column.}
\label{fig:sieve}
\end{figure}

Among various options, sieve trays have been chosen because of their simple, but effective layout, since they have no moving parts, can be easily cleaned, and can be mounted inside the column without welding. A picture of them is shown in Figure~\ref{fig:sieve}.

\subsubsection{Layout of the full-scale plant}
\label{Ch Dist_layout}

\begin{figure*}[t]
\centering
\includegraphics[height=10.5cm]{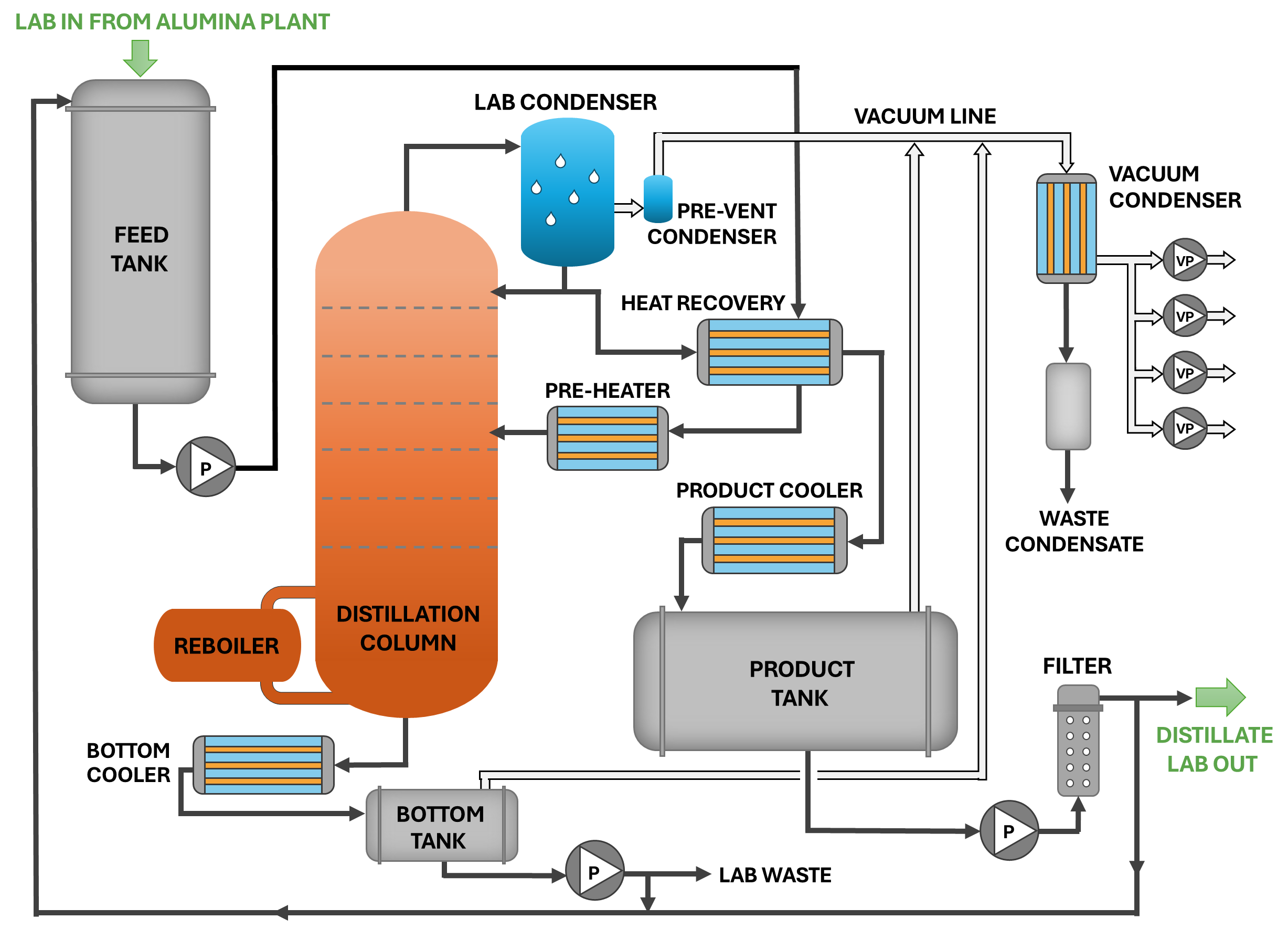}     
\caption{Flowchart scheme of the distillation plant (not in scale). The input LAB is fed into the feed tank, heated up to 180°C by the heat recovery and the pre-heater and inserted in the column above the 3\textsuperscript{rd} tray, where it falls down by gravity in the bottom part. Here it’s evaporated by the reboiler at about 210-220°C. The stream of purified LAB vapours is extracted and condensed in the top part by the condenser and a certain fraction (up to 50$\%$) is sent back to the column as an internal reflux. The distilled LAB is then cooled down to ambient temperature by the heat recovery and the product cooler and stored into the 20 m\textsuperscript{3} horizontal tank. The bottom of the column is discarded regularly into the waste bottom tank to remove the contaminants. The pressure inside the distillation column, the product tank and the bottom tank is kept constant at 5 mbar with a set of 4 vacuum pumps (VP). Finally, after being filtered, the distilled LAB can be either pumped out toward the Mixing plant or sent back to the feed tank for internal loop circulation.}
\label{fig:sketchDist}
\end{figure*}

The plant is designed to operate at 7 m\textsuperscript{3}/h of nominal flow rate. A sketch of its flow diagram is reported in Figure~\ref{fig:sketchDist}, while the main parameters are listed in Table~\ref{tab:dist_parameters}. The LAB entering the plant is collected in a 20 m\textsuperscript{3} vertical feed tank and pumped to the distillation column to start the purification. The liquid is fed at mid-height, above the third tray, after being pre-heated by two counter-flow heat exchangers at about 180°C, thus avoiding destabilizations of the column temperature profile. In the first exchanger, the inlet cold LAB is heated-up by the distilled hot condensate, allowing to recover heat and save energy (in the order of 400 kW\textsubscript{th}); the second one, supplied with hot oil, makes the final temperature adjustments of the column feed. The liquid, falling by gravity, is collected in the bottom part of the column vessel and boiled by the tube-bundle reboiler, thus generating LAB vapours. High-boiling and low-volatility impurities that accumulate in the unevaporated liquid phase remain in the bottom part and are regularly discarded in a waste bottom tank every 30 minutes, up to 2$\%$ of the plant nominal flow rate (7 m\textsuperscript{3}/h).

\begin{table}[htbp]
    \centering
    \caption{Main features and operating parameters of the distillation plant.}
    \begin{tabular}{l|l}
        \toprule
        \textbf{Distillation Plant Parameter} & \textbf{Value} \\
        \midrule
        Sieve Trays & 6 trays \\
        Tray Holes & 3500 holes (diameter 12 mm) \\
        LAB Nominal Flow Rate & 7 m$^3$/h \\
        Reflux Rate & Up to 50\% \\
        Bottom Discharge & 1-2\% \\
        Pressure (at Column Top) & 5 mbar \\
        Temperature @ Reboiler & 210-220°C \\
        Heating Thermal Power & 1000 kW$_{th}$ \\
        Heat Recovery & 400 kW$_{th}$ \\
        Column Diameter & 2000 mm \\
        Column Height & 7 m \\
        Plant Dimensions & 10 m x 9 m x 14 m \\
        Approx. Plant Weight & 55 tons \\
        \bottomrule
    \end{tabular}
    \label{tab:dist_parameters}
\end{table}

With the experience of the pilot plant, the number and type of sensors used to monitor the level of liquid inside the column have been increased, for redundancy: a safe and efficient operation requires the level to be kept within a specific range, to avoid both flooding phenomena or lowering the level below the reboiler. Three level switches, one differential pressure transmitter and one guided-wave level radar are installed in the lower part of the column, while one differential pressure transmitter measures the layers of liquid on the trays.

Purified LAB vapours, that rise up towards the column head, are extracted and liquified by a two-phase condenser, supplied with water at ambient temperature from a cooling-tower. A fraction, up to 50$\%$, of the distilled liquid stream is immediately reintroduced into the column as reflux ratio, hence further increasing the purification efficiency. The remaining part is conveyed by gravity to the 20 m\textsuperscript{3} product tank, after being cooled down to ambient temperature. For technical needs and plant size optimization, the product tank is arranged horizontally and is kept at the same pressure of the column head. This forced us to place the circulation pump of LAB product inside a 4,5 m-deep well, dug into the ground floor, to ensure a suitable NPSH (Net Positive Suction Head) at the pump inlet nozzle, to avoid running the pump dry and to prevent cavitation phenomena. The pump pushes the purified liquid through a set of 50 nm filters, with the purpose of retaining any dust or microparticles that may have been washed off the surfaces.

\begin{figure}
\centering
\includegraphics[height=8.5cm]{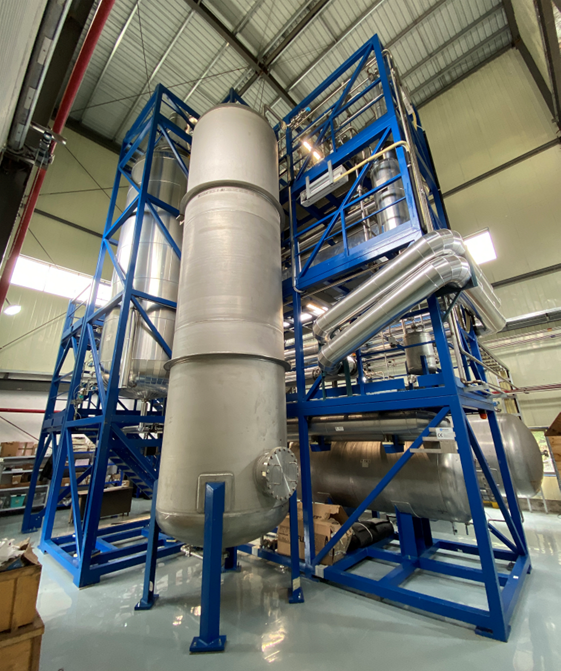}     
\caption{Distillation plant installed in the Over Ground LS building at JUNO site.}
\label{fig:foto_dist}
\end{figure}

Finally, the distilled LAB can be either sent forward to the Mixing plant or circulated back to the inlet feed tank for further purification in internal loop mode.

The plant is equipped with 4 vacuum roots pumps, to keep a constant pressure of 5 mbar at the column head and inside the product and the bottom tanks. The VPs are preceded by a vent condenser, which removes any LAB droplet from the incondensable gases sucked by the pumps. Since in the pilot plant it has been observed that some LAB vapours could anyway reach the VP, especially during the start-up phase, in the full-scale plant a small pre-vent condenser has been added on the vacuum line directly near the column.

Each equipment is supplied with a continuous stream of nitrogen (N\textsubscript{2}) as blanketing to avoid any oxidation.

The distillation plant is now installed in the Over Ground LS building at JUNO site (Figure~\ref{fig:foto_dist}). It has been built and pre-assembled in Polaris’s workshop in 6 skids, plus one vertical and one horizontal tanks, and then shipped to China by sea. Due to its huge dimensions, the installation has been performed from the roof of the building with a 120-ton truck crane, implying challenging engineering solutions.

\subsection{Stripping plant}
\label{Ch Strip}
The stripping plant is the final stage of the purification procedure for JUNO LS. Gas stripping process is used to remove gaseous impurities naturally dissolved into the scintillator, mainly \textsuperscript{222}Rn, \textsuperscript{85}Kr, \textsuperscript{39}Ar radioisotopes, which could generate undesired signals, and O\textsubscript{2}, that is responsible for photon quenching and oxidation in the LS.

\subsubsection{Working principle of the stripping column}
\label{Ch Strip_work}
It is a gas-liquid separation process where gases dissolved in a liquid phase are extracted by desorption mechanisms exploiting a stream of pure stripping gas. This purification technique relies on Henry’s law, which states the dependency of the amount of a gas \emph{i} in a liquid being proportional to its partial pressure (i.e. left side of equation~\ref{Eq:Henry}):

\begin{equation}
    \centering
		  y_i \cdot p_t=K_{H,i} \cdot x_{i}
            \label{Eq:Henry}
\end{equation}

$x_i$ and $y_i$ are the molar fraction of $i$ in liquid and gas phases respectively, $p_t$ the total pressure and $K_H$ the Henry’s constant in atm units. The latter depends on temperature, hence changing this operating parameter can affect the purification efficiency.

The process is based on mass transfer of the contaminant $i$ in liquid phase passing to the vapour phase, that is the stripping gas. In an $x_i$-$y_i$ graph, the flow rates of gas ($G$) and liquid ($L$) streams determine the process operating line, with slope $L/G$, and the so-called stripping factor $S$:

\begin{equation}
    \centering
		  S=\frac{L \cdot K_{H}}{G \cdot p_t}
            \label{Eq:S_factor}
\end{equation}

This quantity represents the removal rate of the contaminant between two equilibrium stages.

\begin{figure}
\centering
\includegraphics[height=6cm]{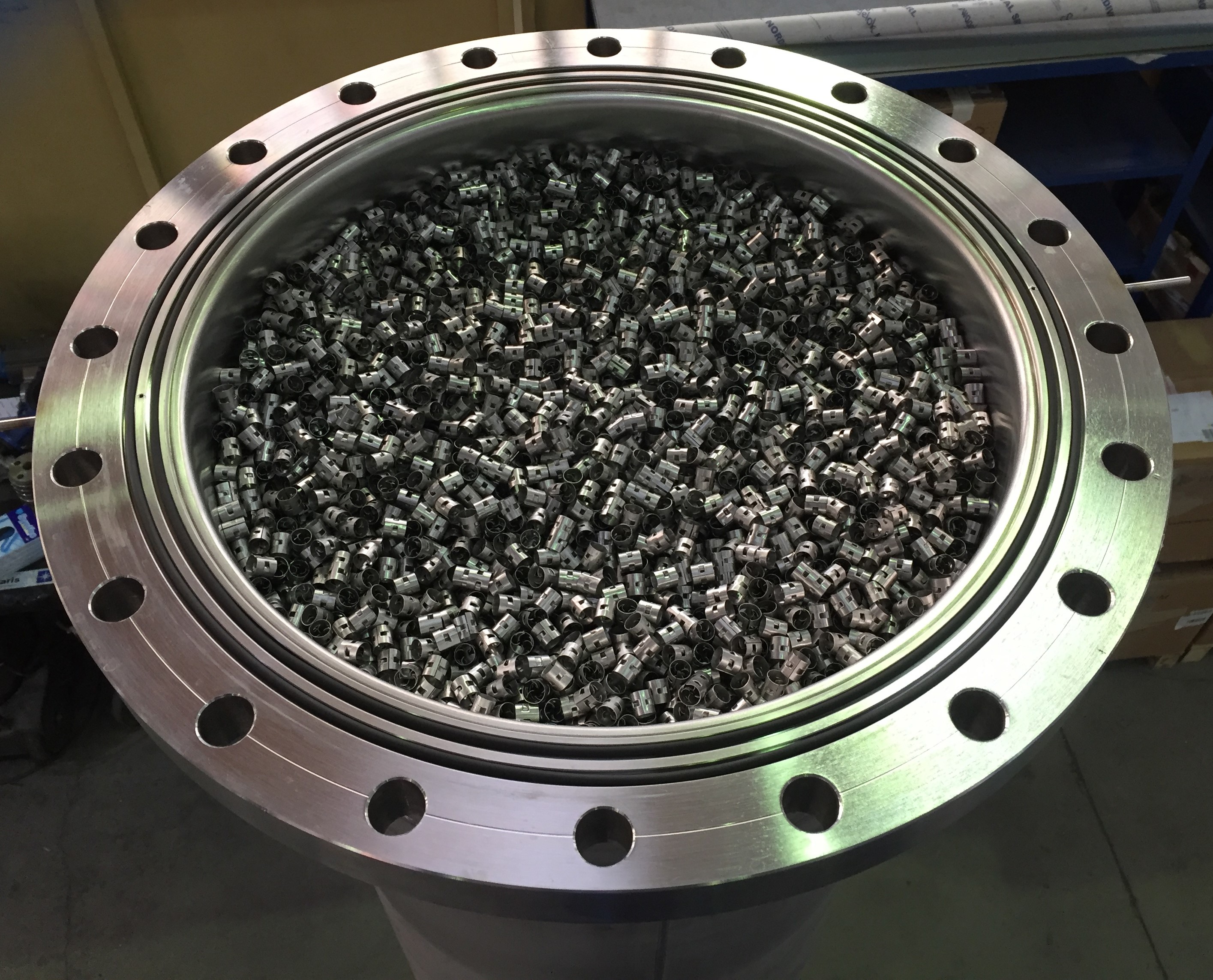}     
\caption{Filling of the stripping column with unstructured packing (Pall rings, 13 mm of diameter).}
\label{fig:Pall_rings}
\end{figure}

Our plant is equipped with a 9 m-high custom-made vertical stripping column filled with unstructured packing, i.e. AISI316 metallic Pall rings of 13 mm diameter (specific interface area $a$ = 430 m\textsuperscript{2}/m\textsuperscript{3}; see Figure~\ref{fig:Pall_rings}), with the purpose of spreading the liquid in thin films or drops, thus increasing the contact surface between the phases. The mass transfer directly depends upon the interface surface exposed between gas and liquid phases: the larger this parameter, the higher will be the purification efficiency. Ultrasonic bath cleaning was adopted to cleanse the Pall rings before assembling and filling the column.

The process is performed in counter-current flow mode, feeding the LS from the top and the gas from the bottom. An adjustable mixture of high-purity nitrogen (HP-N\textsubscript{2}, up to 50 Nm\textsuperscript{3}/h) and ultra-pure water steam (UPW, up to 30 kg/h) is supplied as stripping gas. At top and middle height, the column features a distributor tray, which restores a uniform spatial distribution of liquid flux inside.

Also this plant is designed to operate in partial vacuum, at about 250-300 mbar, and up to 90°C, so that the LS viscosity is reduced and the global efficiency is increased, as can be inferred from equation~\ref{Eq:S_factor}.

In the Daya Bay tests, the stripping process was performed at 90°C at 300 mbar and had a purification efficiency for \textsuperscript{222}Rn of 95,8$\%$ on a 115 L/h LS flux with a stream of 1 Nm\textsuperscript{3}/h HP-N\textsubscript{2}. The dimensions of the column have been scaled up to design the final plant for JUNO, but the height has been further increased of 50$\%$. Based on simulations, the efficiency for the full-scale plant is foreseen to be about 99$\%$ for a LS flow rate of 7000 L/h, with 30 Nm\textsuperscript{3}/h HP-N\textsubscript{2} as stripping gas.

\subsubsection{Layout of the full-scale plant}
\label{Ch Strip_layout}
\begin{figure*}[t]
\centering
\includegraphics[height=10.5cm]{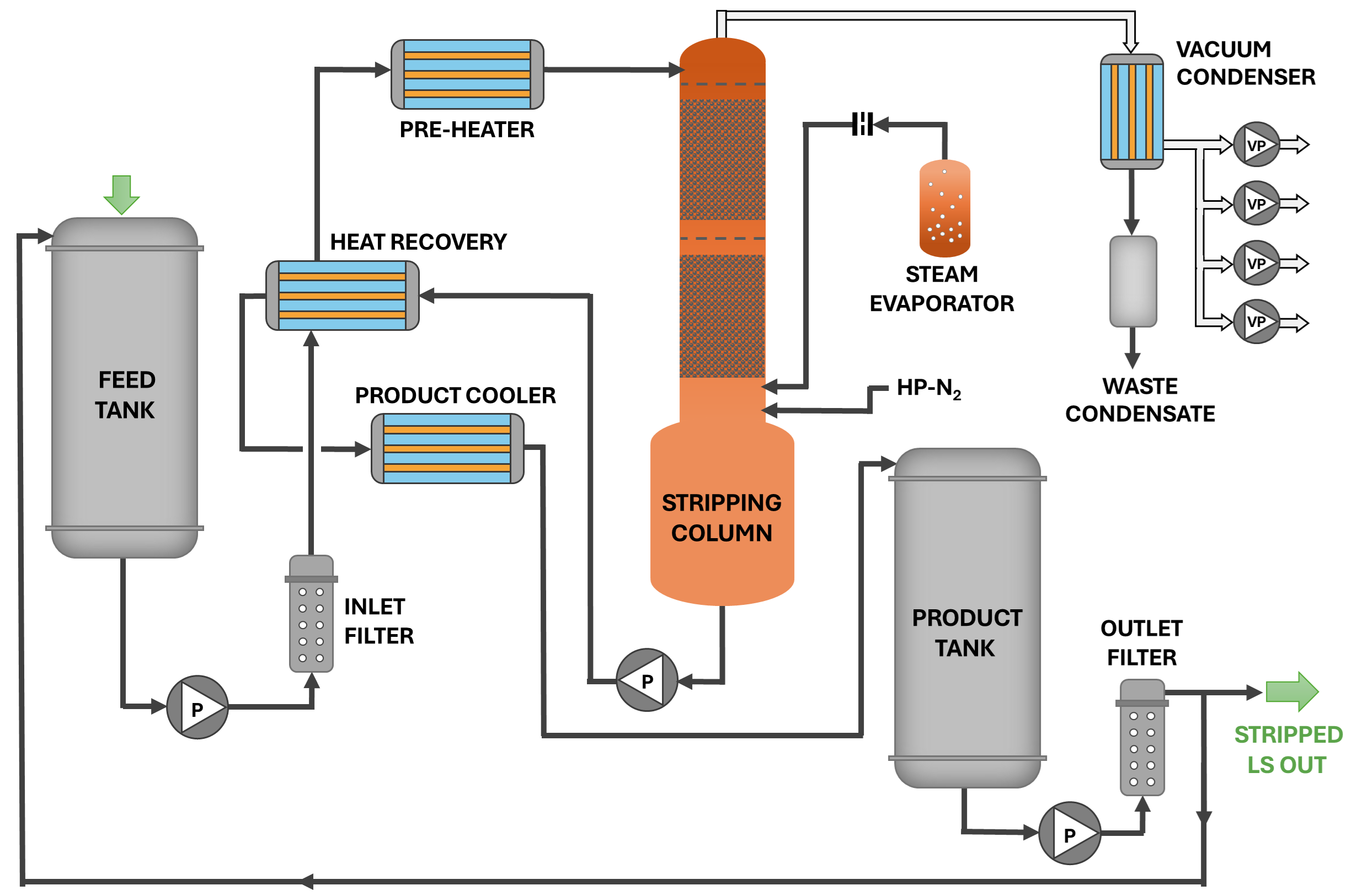}     
\caption{Flowchart scheme of the stripping plant (not in scale). The LS entering the plant is pumped from the 20 m\textsuperscript{3} feed tank through the inlet filters and heated by the heat recovery and the latter heat exchanger. The LS is then inserted into the stripping column from the top, falling down by gravity, while the stripping gas, an adjustable mixture of high-pure nitrogen and ultrapure water steam, is fed from the bottom. The purified LS is collected at the bottom of the column, cooled down to 21°C and stored into the product tank. Finally, it is filtered passing through the outlet filters and it can be pumped either to the next stage or sent back to the feed tank for internal loop circulation. The vacuum level inside the column is kept constant by a system composed of a vacuum condenser and 4 vacuum pumps.}
\label{fig:sketchStrip}
\end{figure*}

The design and layout of the plant are based on a nominal LS flow rate of 7 m\textsuperscript{3}/h. The flow diagram and the operating parameters are reported in Figure~\ref{fig:sketchStrip} and Table~\ref{tab:strip_parameters}, respectively.

The LS, delivered by the Water Extraction plant, is collected into the 20 m\textsuperscript{3} vertical feed tank and pumped through a first set of 50 nm filters, to prevent dust from previous stages or transfer pipelines to pollute the internal surfaces and the column packing. The scintillator is heated-up to the process temperature, before being fed at the column head. Similarly to the distillation plant, we take advantage of heat recovery from stripped hot scintillator in a first heat exchanger, while a second one adjust the temperature to the target value.

\begin{table}[htbp]
    \centering
    \caption{Main features and operating parameters of the stripping plant.}
    \begin{tabular}{l|l}
        \toprule
        \textbf{Stripping Plant Parameter} & \textbf{Value} \\
        \midrule
        Column filling & Unstructured (Pall-rings) \\
        Pall-Rings diameter & 13 mm \\
        Specific Interface Area & 430 m$^2$/m$^3$ \\
        LS Nominal Flow Rate & 7 m$^3$/h \\
        HP N$_2$ Flow Rate & Up to 50 Nm$^3$/h \\
        UPW Steam Flow Rate & Up to 30 kg/h \\
        Pressure & 250 mbar \\
        Temperature & 70-90°C \\
        Heating Thermal Power & 200 kW$_{th}$ \\
        Heat Recovery & 160 kW$_{th}$ \\
        Column Diameter & 500 mm \\
        Column Height & 9 m (5.6 m of packing) \\
        Plant Dimensions & 6.5 m $\times$ 9 m $\times$ 12 m \\
        Approx. Plant Weight & 35 tons \\
        \bottomrule
    \end{tabular}
    \label{tab:strip_parameters}
\end{table}

The falling liquid is contacted and stripped by the upward stream of gas mixture. The HP-N\textsubscript{2} is supplied by a dedicated plant, exploiting purification through low-temperature adsorption technology to remove radioisotopes from the gas. This is important because their residual content sets the purification limit that can be reached in the stripping process.

The water steam (30 kg/h maximum) is produced at 65°C at about 250 mbar by an oil-based evaporator, starting from ultra-pure water (\textgreater18 MΩ$\cdot$cm). A calibrated orifice of 1 cm in diameter controls the flux injected into the column. As soon as the vapors enter the column, which is kept at higher temperature, they become superheated, hence preventing condensation.

\begin{figure}
\centering
\includegraphics[height=9cm]{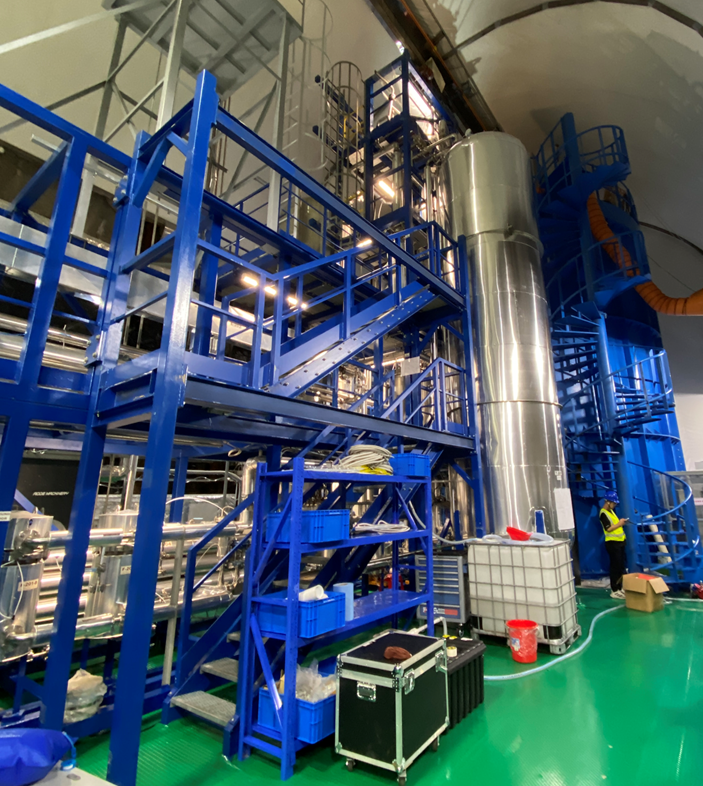}     
\caption{Stripping plant installed in the Underground LS Hall at JUNO site.}
\label{fig:foto_strip}
\end{figure}

After being stripped, the LS is cooled down to 21°C and stored into the 20 m\textsuperscript{3} vertical product tank. The final step is the filtration through another set of 50 nm filters, that restrain any dust or metallic particulate that could have been release by the Pall rings or other surfaces. The purified LS can be either sent to the Central Detector of JUNO for its filling procedure or recirculated back to the inlet feed tank for internal operation, in case of need.

A set of 4 vacuum roots pumps is connected at the column head to keep the pressure constant at 250 mbar and suck the exhausted stripping gas out of the column. Condensable vapours, as water steam, are liquified and drained by a vent condenser, while the non-condensable ones, like nitrogen, are expelled by the pumps themselves.

The stripping plant is installed in the Underground LS Hall at JUNO site. It is composed of 2 vertical tanks and 3 skids, with 5 walkable floors in total. Each unit has been transported to the underground laboratory though the 1,5 km slope tunnel and assembled using the overhead crane installed on the LS Hall roof. A picture of the stripping plant is shown in Figure~\ref{fig:foto_strip}.

\subsection{Common Features}
\label{Ch Common_F}
Some common criteria and concepts have driven the design and the production of these two plants.

\subsubsection{Cleanliness and air tightness requirements}
\label{Ch Cleanliness}
Several additional precautions have been adopted in order to achieve the desired LS purity and avoid spoiling it.

Selected 316L stainless steel was chosen as material to manufacture all the main components of the plants and all parts contacted with the scintillator. In addition, a special cleaning procedure was studied and implemented to treat the surfaces and prevent oxidation phenomena: a mirror finishing of the steel through electropolishing below 0,8 $\mu$m; pickling and degreasing of the welds; a mild passivation to create an inert layer, using a solution of nitric acid (25$\%$) and demineralized water (75$\%$); rinsing with water until conductivity and pH values were restored; compressed air-drying process with dehumidifier and coalescing filters and final storage under N\textsubscript{2} atmosphere. Several quality controls were performed to certify the cleaning procedure, such as particle counting test, endoscopic inspection, and white wipe test. A final rinsing with high purity water was performed at JUNO site, after plant installation and skids assembling. The rinsing water has been analysed via particle counting to fulfil MIL STD 1246C – Class \textless 50 cleanliness level~\cite{MIL}.

To maintain the required level also during run phase, 50 nm pore size pre-wetted filters have been installed on the scintillator pipelines, with the aim of preventing circulation of dust and metallic particulate that could be present inside the plants, despite the special cleaning procedure adopted. To allow maintenance operations on filters without forcing a stop of the plant, every filtering stage is composed by two independent filter holders, which allocates six 20 inch-filter cartridges each. In the distillation plant, the filtration is carried out as final step on the outlet line, while for the stripping both on the inlet (before the stripping column, to avoid the dirty being trapped into the unstructured packing) and outlet lines.

Since radioactive gases and oxygen can spoil the LS, all the plant must be air-tight proof. The layout of the flanges has been designed with a double o-ring protection (Viton gaskets for flange exposed to temperatures \textless150°C, Kalrez gaskets for higher temperatures). Each component, flange or moving part have been tested with a helium leak detector to check the sealing. JUNO requirements set the thresholds of \textless10\textsuperscript{-8} mbar·L/s for single leak rate and \textless10\textsuperscript{-6} mbar·L/s as overall integral leak rate. Everything was certified to be within specifics. Moreover, the flanges are continuously purged by a constant flux of HP-N\textsubscript{2}, to further reduce any possible air leakage from outside.

A procedure of vacuum pumping down to mbar level and nitrogen purging above 1,2 bar was repeated three times, to remove air before filling the plant with the scintillator. During operation phase, all tanks and the columns are kept under HP-N\textsubscript{2} blanketing, in overpressure (compared to atmospheric pressure) whenever feasible.

\subsubsection{Mechanical aspects}
\label{Ch Mechanics}
All main components and pipelines have been pre-assembled and mounted inside skids before shipping them to China. Orbital welding or TIG welding were adopted whenever possible, while interconnections between skids or equipment have been fitted with flanges sealed with o-rings. For ¼” sampling ports, VCR fittings with metal gaskets were chosen.

All main pipelines for LS circulation have been equipped with full-bore ball valves, while pneumatic regulating valves have been installed in crucial points to remotely control the most important plant parameters, such as feed and product fluxes and the pressure in the column.

To improve the thermal efficiency, Rockwool is adopted as insulating material for hot pipelines and equipment, while Armaflex for the cold ones.

Each plant is provided with 4 vacuum multi-stage roots pumps, mounted in parallel and located in the upper floors of the plant, near the main column, to keep the design vacuum level during the purification process. The 3 magnetic driven centrifugal pumps (CDR pompe s.r.l. company) for the circulation of the scintillator inside the plant are positioned on the ground floor (or even beneath it, in the distillation plant) to ensure stability and a proper NPSH. 

Both plants can be operated either in internal loop mode, during self-cleaning, commissioning and start-up operations, where the liquid is recirculated back to the input feed tank, or in production mode, where the processed scintillator is sent forward to the next stage of the purification sequence.

\subsubsection{DCS and automation}
\label{Ch DCS}
The plants are controlled by a reliable and safe Distributed Control System (DCS), which sets the operating parameters, monitors their trend, and actuates an alarm $\&$ interlock system in case of out-of-range values or emergency.

It is based on a Siemens PLC, model ET200S, installed in a control panel along with the I/O modules, and it is supervised by a local pc with a dedicated SCADA software with user interface (UI). The PLC communicates with the pc through a MODBUS serial line, while field signals from instruments and sensors are transmitted to control panel though 4-20 mA wiring protocol.

The PLC has three main logic areas: the main control functions, always active to monitor the operating parameters and adjust accordingly the automatic valves and items to reach the setpoint value; the automatic sequences, that are set by the operator to be run independently and automatically by the system under certain conditions; the alarms and interlocks, which set the operating and safety ranges for each variable, provide acoustic and visual alarms if the ranges are exceeded and inhibit some functionalities in case of emergency, until shutdown in the worst cases. The interlocks can be software based, the most common one, where the inhibition is activated by the SCADA software with logic commands (but can also be bypassed in case of need), or hard-wired based, used only for critical items, where the electric signal is physically interrupted by opening the circuit.

The supervision system relies on a user-friendly panel with different pages, that allows an easy and efficient monitoring of the plant. The Run page displays a scheme of the plant, with all the field variables measured in real-time. The second page collects a list of all alarm signals that have occurred, both past and ongoing. In the last page, several graphical trends for the main operating parameters are reported: they show the history of values measured during time for each variable and they are crucial to understand the dynamic behaviour of the plant.

\subsubsection{Safety}
\label{Ch Safety}
The plants have been studied to fulfil several international regulations and certifications and they have received the EC declaration of conformity.

Skid frames and structural loading diagrams have been designed to meet both European and Chinese safety regulations. For JUNO experimental site, all equipment must comply with the National Standard of People’s Republic of China Code of Seismic Design of Buildings GB 50011-2001. During calculations, grade VII seismic fortification intensity with a 0.10g horizontal acceleration was considered. 

All tanks, vessel, columns, and main components have been certified according to PED - Group1 – Category IV – Module G classification of PED directive for pressurized equipment. After the construction phase, they have undergone both overpressure and vacuum tests. Rupture disks are installed on all tanks and the two columns and are designed to break at 3,5 bar\textsubscript{g}, while a pressure safety valve is mounted on the HP-N\textsubscript{2} feed line. The plants have also been licensed by the Chinese authority SELO (China Special Equipment Licensing Office).

For fire protection, all electrical equipment is provided with certification for ATEX area classification, in Class 1 Zone 2 T2. Actually, in the distillation plant, the LAB is heat-up to temperatures above its flash point.

Security lights, active also in case of power outage, have been installed to illuminate any access points and paths.

A detailed Hazop (HAZard and OPerability) analysis for industrial risk assessment was carried out in collaboration with a team of international experts. The major hazards arising from normal and abnormal operation were identified, together with their occurrence probability and severity. Each functional block of the plant was isolated and examined for possible deviations from the design intention. For the hazards, preventive and remedying actions were defined.

\section{Commissioning and preliminary results}
\label{Ch Commissioning}
During an intensive commissioning phase, the distillation and stripping plants have been run repeatedly, first in internal loop mode to test and optimize the operating parameters, then in production mode for the joint commissioning with all the other purification plants. They were operated for more than 8h uninterruptedly for many times, in order to test both the stability and the repeatability of the operating conditions. All parameters were tuned to maintain a production flow rate of 7 m\textsuperscript{3}/h, which is required to complete the CD filling in 6 months, and all other nominal values, while providing an efficient purification. Both demonstrated an excellent stability and safety level throughout the whole testing period.

For the distillation plant, the condenser for purified LAB vapours after the distillation process represents the most critical component to be set, since the condensed liquid is split between the reflux stream back to the column and the product one, send to the production tank, that must be kept at nominal 7 m\textsuperscript{3}/h. Fine adjustments must be done to control both the flow rates and the outlet temperature from the condenser. Subcooling below 100°C should be avoided to prevent thermal destabilizations of the column by the reflux and take advantage of the heat recovery exchanger exploiting the product stream. For these reasons, the condenser’s cooling power was gradually tuned by carefully regulating its valves, while continuously monitoring the LAB outlet values. The optimal configuration was found to be about 185°C at the column top, 130°C as outlet temperature from the condenser and a reflux flow rate of 4-5 m\textsuperscript{3}. During the whole commissioning, almost 250 m\textsuperscript{3} have been distilled in total, in order to tune all working parameters and gain experience in managing the plant during all phases, from the start-up to the shutdown.

Concerning the stripping plant, several tests of the stripping purification process were carried out using both HP-N\textsubscript{2} and UP water steam, to check the feasibility. With the latter, some issues emerged because of water solubility levels into LAB changing with temperature. Actually, cooling the scintillator back to room temperature after stripping at higher temperatures caused dissolved water to condense, resulting in a milky emulsion of suspended water droplets, that temporarily compromised the LS transparency. Following a recovery period exceeding 24 hours, the LS and water separated, thereby restoring the quality of the first. Nonetheless, it was determined safer to exclusively use nitrogen for the stripping process, to prevent an excessive water content in the scintillator.

After this effect was pointed out, further investigation also on the water extraction process (refer to section~\ref{Ch purif_strategy}) was performed through a joint commissioning together with the stripping plant. In particular, the removal efficiency of water and the residual water content after the stripping process at different process temperatures and pressures were tested. The temperature was ranged between 70°C and 90°C, while the pressure inside the column from 250 mbar to 350 mbar. The water content essay was performed before and after the stripping process with a dedicated apparatus. The best working configuration was determined to be 70°C at 250 mbar, using 15 Nm\textsuperscript{3}/h HP-N\textsubscript{2} as stripping gas: with these parameters, the water content was reduced from 154 ppm down to 20 ppm, which complies with JUNO standards.

Concerning optical and radiopurity features, to carefully verify the performances and the purification efficiencies of both plants, a comprehensive set of several measurements with different techniques was established, including absorption and emission spectra, particle counting, evaluation of the attenuation length, ICP-MS and NAA. After each plant operation, samples were taken and analysed to check compliance with the stringent JUNO requirements. Preliminary results of some of them are presented in the following.

\begin{figure}
\centering
\includegraphics[height=7cm]{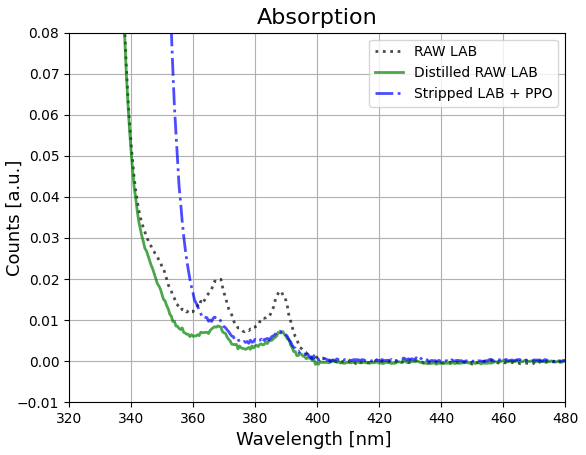}     
\caption{Absorption spectra, in arbitrary units a.u., of raw LAB (grey, dotted line), distilled LAB (green, solid line) and stripped LAB+PPO (blue, dash-dot line).}
\label{fig:abs}
\end{figure}

In Figure~\ref{fig:abs}, is reported a comparison between the absorption spectra, in arbitrary units a.u., of raw LAB (grey, dotted line), distilled LAB (green, solid line) and stripped LAB+PPO (blue, dash-dot line). The absorption peaks at 370 nm and 390 nm of the grey curve have been crucially decreased after the distillation process performed directly on raw LAB, thus proving also good optical purification capabilities of this plant. Then, the solvent has been mixed with PPO, sent to underground and purified by the stripping process, resulting in the blue line: the insertion of PPO causes a natural shift of the absorption edge toward longer wavelengths, around 355 nm, but all the rest of the line is almost overlapped to the green one, meaning that this process did not affect nor introduce major optical contamination sources into the scintillator, also thanks to the cleaning and sealing solutions adopted.

This was confirmed also by the excellent particle counting results, which are reported in Table~\ref{tab:ParticleStr}. These values should be compared with the ones shown in Table~\ref{tab:Class50}, i.e. the Class 50 level of the MIL STD 1246C~\cite{MIL}, adopted as standard reference in JUNO. For each particle size, the number of particles present in the sample must not exceed the threshold value. In Table~\ref{tab:ParticleStr}, the result of the tests on samples collected right after the stripping plant. The sensitivity of the instrument was much higher than required by the Class 50 standard; anyway, the counts for the largest particles detected (0,3 and 0,5 $\mu$m) in the samples were already 0, while for Class 50 requirement the smallest particle size to be monitored is 5 $\mu$m and the counts should be \textless1660. This clearly demonstrates an excellent cleanliness level of all main components and an efficient filtration system.

\begin{table}[ht]
    \centering
    \caption{The MIL STD 1246C – Class 50 cleanliness level, adopted as standard reference for JUNO LS: the number of particles must be lower than the threshold value for each particle size.}
    \begin{tabular}{c|c}
        \toprule
        \multicolumn{2}{c}{JUNO Class 50  (from MIL 1246C)}\\
        \midrule
        Particle size ($\mu$m)& Count/L \\
        \midrule
        5 & 1660 \\
        15 & 250 \\
        25 & 73 \\
        50 & 10 \\
        \bottomrule
    \end{tabular}
    \label{tab:Class50}
\end{table}

\begin{table}[ht]
    \centering
    \caption{The result of the particle counting test performed on LS samples collected after the purification through stripping plant.}
    \begin{tabular}{c|c}
        \toprule
        \multicolumn{2}{c}{Particle counting on stripping plant samples}\\
        \midrule
        Particle size ($\mu$m)& Count/L \\
        \midrule
        0,1 & 1200 \\
        0,15 & 500 \\
        0,2 & 100 \\
        0,3 & 0 \\
        0,5 & 0 \\
        \bottomrule
    \end{tabular}
    \label{tab:ParticleStr}
\end{table}

Others QA/QC measurements are still ongoing to fully characterize the purified samples after each step of the purification sequence, exploiting advanced techniques and pushing their sensitivity and detection limit to values beyond the state of the art. After a long phase of R$\&$D, they have been optimized, refined, and validated. Their preliminary results on several purified LS samples seem very promising in terms of LS quality and purification and they will be published soon.

\section{Conclusion}
\label{Ch Conclusion}
This paper summarises the design, main features and operation of the distillation and stripping purification plants, which have been built and installed at the JUNO site for the purification of the LAB-based liquid scintillator for JUNO experiment. Thanks also to a preliminary test campaign at Daya Bay laboratories, both these processes have been proven crucial for the accomplishment of the optical and radiopurity JUNO requirements, which are mandatory for its scientific program. The layout and design of the plants were driven by precise technical needs, including low background, cleanliness level, leak tightness and safety.

The plants have been recently commissioned and successfully tested to prepare the 6-months filling phase of the JUNO detector. Preliminary results show very effective purification performances and stable operating conditions of the plants.

\section*{Acknowledgments}
\label{Ch Ack}
This work is funded by the Istituto Nazionale di Fisica Nucleare (INFN), Italy, through the JUNO experiment. Special thanks go to Michele Montuschi, for his key contributions to the development of the DCS control system; Mario Masetto, Eleonora Canesi, Gabriele Milone and the Polaris staff for their work on the pilot and full-scale purification plants; and to the Chinese colleagues Hu Tao, Zhou Li, Yu Boxiang, Cai Xiao, Fang Jian, Sun Lijun, Sun Xilei, Xie Yuguang, Ling Xin, Han Hechong and Huang Jinhao from IHEP for their invaluable help and support during the plant operation onsite.

\bibliographystyle{elsarticle-num}
\bibliography{example.bib}









\end{document}